# A multiscale Bayesian approach to quantification and denoising of energy-dispersive x-ray data

Pau Torruella[1], Abderrahim Halimi[2], Ludovica Tovaglieri[3], Céline Lichtensteiger[3], Duncan T. L. Alexander[1], Cécile Hébert[1]

*1 Electron Spectrometry and Microscopy Laboratory (LSME), Institute of Physics (IPHYS), Ecole Polytechnique Fédérale de Lausanne (EPFL), Lausanne, Switzerland.*

*2 School of Engineering and Physical Sciences, Heriot-Watt University, Edinburgh UK.*

*3 Department of Quantum Matter Physics, University of Geneva, Geneva, Switzerland.*

## Abstract

Energy dispersive X-ray (EDX) spectrum imaging yields compositional information with a spatial resolution down to the atomic level. However, experimental limitations often produce extremely sparse and noisy EDX spectra. Under such conditions, every detected X-ray must be leveraged to obtain the maximum possible amount of information about the sample. To this end, we introduce a robust multiscale Bayesian approach that accounts for the Poisson statistics in the EDX data and leverages their underlying spatial correlations. This is combined with EDX spectral simulation (elemental contributions and Bremsstrahlung background) into a Bayesian estimation strategy. When tested using simulated datasets, the chemical maps obtained with this approach are more accurate and preserve a higher spatial resolution than those obtained by standard methods. These properties translate to experimental datasets, where the method enhances the atomic resolution chemical maps of a canonical tetragonal ferroelectric $PbTiO_3$ sample, such that ferroelectric domains are mapped with unit-cell resolution.

## Introduction

Energy dispersive X-ray (EDX) spectroscopy is a technique that can provide chemical information with sub-nanometric resolution when performed in a scanning transmission electron microscope (STEM)[1]. In the measurement, a high-energy electron beam is focused to a small probe that ionizes the atoms at a particular spatial position in a sample of interest. As the sample's atoms deexcite through various processes, they emit X-rays with a certain probability. This emission occurs at different characteristic energies (X-ray "lines") that depend on which chemical elements are present in the sample at the probe position. The histogram of X-rays emitted at each energy, i.e., the EDX spectrum, constitutes a signature of the chemical composition. By raster scanning the electron beam over the sample, a spectrum image (SI) can be acquired, containing a full EDX spectrum at each spatial pixel. Through spectral analysis, the atomic percentages of each element present at that position can be estimated, enabling quantified chemical mapping. Commonly, this quantification is done through the well-established Cliff-Lorimer approach[2], or the more sophisticated $\zeta$-factor method[3]. These involve integrating the signal above a spectral background (mainly originating from Bremsstrahlung radiation) for given spectral lines and computing their intensity ratio in conjunction with application of empirical or calculated constants. There are, however, several factors that can pose a challenge to this approach.

First, the limited collection efficiency of EDX spectrometers is a prominent issue. The development of compact silicon drift detectors (SDDs) with flexible geometry, that can be placed within the optical pole piece of the microscope in close proximity to the sample, greatly improved collection efficiency compared to the previous SiLi detectors through increased X-ray collection solid angle. For instance, the 4 SSD segment ChemiSTEM system used here has a total collection solid angle of almost 1 Sr, and commercialized systems have reached collection solid angles of >4 Sr [4]. Nevertheless, even in this case, most of the generated X-rays do not reach the detector. Absorption in the sample, shadowing effects and dead time in SDDs are additional factors that hinder X-ray detection[5].

Another limitation is the beam-induced degradation of the sample. Depending on its nature and the acceleration voltage used in the STEM, knock-on damage, radiolysis, and heating can alter the material that is being studied.[6,7] This in turn limits the maximum acquisition time and beam current/beam flux that can be used for an experiment. Additionally, sample contamination, usually in the form of carbon deposition, can further limit the total X-ray signal that is collected[8,9].

Consequently, the acquired SI spectra usually have very poor statistics. While, ideally, a single pixel spectrum would contain thousands of X-ray counts, in practice, only a few tens are collected (figure 1A). In these instances, we must use advanced analysis methods to make the most out of our data. Multivariate analysis[10,11] (MVA) and deep learning[12–14] have all been used in different instances in this regard. However, MVA methods such as spectral decomposition inherently make assumptions about the sample, i.e. that the data matrix can be efficiently approximated by a very low rank matrix. This assumption can in turn lead to artefacts such as the introduction of random error biases[15,16]. When it comes to deep learning, one must first train the chosen model. For this, having training data of sufficient quality and abundance is crucial to avoid common machine learning pitfalls such as overfitting, biases, and propagation of imbalances in the training set[17]. However, suitable training data is often not available, or, simply, defeats the purpose of the measurement (for example in the common case where the goal of an experiment is the initial exploration of a system).

In these terms, an ideal analysis is one that gives the most precise elemental quantification possible while avoiding artefacts, and that should be applicable to any sample without prior knowledge. Any mention of "precision" in the quantification should further be accompanied by an equally well-defined estimate of the uncertainty of the measurement. In this work, we propose a strategy that addresses all of these requirements.

To improve the treatment of STEM-EDX data, we consider several factors. First, the physics of the interaction that gives rise to the X-ray signal is well-established and can be built into the statistical analysis. Following this idea, some recent work has addressed the physics-guided spectral decomposition of EDX spectrum images, recovering an accurate representation of the sample chemistry from data with poor statistics[10,18,19]. While this improved the quality of the decomposition, a priori knowledge in terms of the data matrix rank is still needed. Moreover, even if more performant than conventional ones, this decomposition method still requires a minimum amount of signal for a given energy channel in the SI to avoid problems when diagonalizing the data matrix or when performing updates on the solution. In order to address these limitations, here we abandon the decomposition approach in favour of a Bayesian estimation strategy combining data statistics with available prior knowledge to calculate the spectral emission of each chemical element and the Bremsstrahlung background emission, taking into account relevant X-ray detection parameters of the experimental setup.

Second, the spatial correlation of spectra acquired in the sample is considered, as opposed to exclusively exploiting the spectral correlation as done by pure matrix decompositions methods. This is done within the framework of a multiscale approach, whereby the resolution of the original dataset is reduced at different levels, proportionally increasing the signal-to-noise ratio (SNR) of each spectrum, and then performing the Bayesian estimation of the chemical composition at all of these levels. As an added benefit, the uncertainty of the measurement is also directly obtained by the Bayesian approach, which allows the reliability of the quantification to be evaluated. Such an uncertainty estimation is very challenging to deduce in other approaches to EDX quantification, since the uncertainty in the $k$ or $\zeta$ factors, the effect of the discreteness of the signal, and the uncertainties of the background fitting, all have to be propagated.

These steps combine into a robust multiscale Bayesian (RMB) approach that outputs quantified elemental maps in a reliable and denoised manner from STEM-EDX SIs. The method proves successful even when the X-ray emission is sampled very sparsely, and the total number of counts is small.

We finally demonstrate the advantages of our method by analysing atomic-resolution STEM-EDX data of ferroelectric $PbTiO_3$. In a general sense, atomic-resolution SIs are one of the most challenging types of EDX measurements, given that they typically require a sub-Å probe size, increasing the electron beam current density. However, this is counterbalanced by a current density threshold to damage that many materials exhibit[22], in turn limiting the usable probe current and hence absolute X-ray counts obtained. The high spatial resolution and small sample thickness required for such imaging also massively reduce the number of atoms probed at each pixel position, further reducing the number of counts in each pixel spectrum (see figure 1A). Under these conditions, a method that can make the most out of the recorded data is crucial, as we demonstrate using the $PbTiO_3$ SI data. $PbTiO_3$ is a tetragonal ferroelectric with a bulk critical temperature ($T_C$) of 765 K. The ferroelectric polarisation is closely linked to the Ti cation displacement from the centrosymmetric position within the unit cell along the tetragonal c-axis[20]. Therefore, an accurate determination of atomic positions allows the local polarisation to be estimated[21]. Here, by comparing to a standard chemical mapping approach, the proposed RMB analysis is decisive for recovering atomic-resolution chemical maps of enough quality that the local polarisation can be directly measured from them.

## Methods

### *Algorithm description*

The algorithm is based upon ideas originally developed for the reconstruction of 3D LIDAR data[23]. Let us denote by $y_{n,t}$ the detected X-ray counts at the spatial position $n \in \{1, \dots, N\}$ and energy channel $t \in \{1, \dots, T\}$, where $N$ and $T$ represent the total number of pixels and energy channels, respectively. Then, let us refer to the spectral signature of the $k_{th}$ element in the sample scaled by the probability of emission per atom in the sample (its cross-section) as $s_{k,t}$; hereafter, $s_{k,t}$ are referred to as "endmembers". For relatively thin samples, where absorption does not represent a relevant contribution, the X-ray emission probability of each element in the periodic table is simulated through the ESPM library.[19] The generated atomic X-ray signatures account for the incident electron energy (STEM high tension), and the take-off angle and detective quantum efficiency of the X-ray detectors. Using endmembers generated in this manner, the quantification approach inherently adapts to changes in the experimental setup and conditions. The Bremsstrahlung background is also simulated through ESPM by means of additional $s_{k,t}$ endmembers.

Assuming Poisson statistics for the detected X-rays, and statistical independence between the measured spectra, leads to the following expression for the joint likelihood given the chemical composition $R$:

$$P(Y|R) = \prod_{n=1}^{N} \prod_{t=1}^{T} \frac{x_{n,t}^{y_{n,t}}(r_n)}{y_{n,t}!} exp[-x_{n,t}(r_n)] \qquad (1)$$

Where $x_{n,t}(r_n) = \sum_{k=1}^{K} r_{n,k} s_{k,t}$ in which $r_{n,k} \geq 0$ represents the abundance for each element present in the sample and $y_{n,t}!$ denotes the factorial operator. This expression can be approximated (see details in the supplementary information) by:

$$P(y_n|r_n) \propto \prod_{k=1}^{K}[\mathcal{G}(r_{n,k}; 1 + \overline{r_{n,k}}, S_k)\bar{Q}(y_{n,t})] \qquad (2)$$

Where $S_k = \sum_{t=1}^{T} s_{k,t}$, $\mathcal{G}(x, \gamma, \theta) \propto x^{-\gamma} exp\left(-\frac{x}{\theta}\right)$ denotes the gamma distribution, and $\bar{Q}(y_{n,t})$ is a normalization constant that depends on the signal counts (but not in the abundances, *r*). $\overline{r_{n,k}}$ denotes an initial estimate of the abundance at the $n_{th}$ pixel location for the $k_{th}$ element, for which we used the Sunsal unmixing algorithm[24]. This observation model will be used to calculate the elemental abundances through maximum likelihood estimation (MLE).

A common approach to improve the MLE performance for sparse counting data is to consider muti-scale information[25–28]. The main insight is that spectra filtered with a uniform kernel at a low spatial frequency maintain their Poisson distribution but provide lower-noise abundance estimates, though this comes at the expense of decreased spatial resolution. We previously exploited a similar idea for STEM-EDX data in the context of spectral decomposition, where spectral components obtained from downsampled SIs were used to initialize a decomposition for the dataset at full resolution[29]. Here, we adopt a more general strategy by considering *L* low-pass filtered versions of the SI. These are computed based on predefined graphs $\phi^{1,\dots,L}$. The spatially low-pass filtered data are denoted by $Y^l$ for a low pass filtering with a 2*l*-1 by 2*l*-1 uniform window. Assuming independence between the downsampled spectra leads to *L* likelihood distributions:

$$P\left(y_n^{(l)}\middle|r_n^{(l)}\right) \propto \mathcal{G}\left(r_{n,k}^{(l)}; 1 + \bar{r}_{n,k}^{(l)}, 1\right)\bar{Q}\left(y_{n,t}^{(l)}\right) \text{ for } l \in \{1, \dots, L\} \quad (3)$$

Estimating the elemental abundances is challenging when few X-ray photons are detected per spectrum. To overcome this issue, we apply a Bayesian strategy where the approximate likelihood (3), is combined with the prior distributions of the parameters in the model accounting for abundance non-negativity and parameter spatial correlation. The resulting posterior distribution will be exploited by deriving Bayesian point estimators and additional measures of uncertainty about the estimates.

Under our acquisition conditions, it can be assumed that abundances will change smoothly in the sample, translating into a spatial correlation between neighbouring pixels. This "spatial smoothness" can be enforced by introducing an *N x K* latent variable *M* with a Gaussian prior distribution as follows:[30,31]

$$m_{n,k}|r_{v_n,k}^{(l)}, v_{v_n,k}^{(l)}, \psi_{n,k}^2 \sim \prod_{n' \in v_n} \left[\prod_{l=1}^{L} \mathcal{N}\left(m_{n,k}; r_{n',k}^{(l)}, \frac{\psi_{n,k}^2}{v_{n',n,k}^{(l)}}\right)\right] \quad (4)$$

In (4) the subindices *n* and *n'* refer to a given pixel position and a given neighbouring pixel respectively, while $v_n$ refers to the elements in the considered neighbourhood of pixel *n*. The parameter $\psi_{n,k}^2$ represents the variance of the latent variable and contains the abundance uncertainty information of the $k_{th}$ element. $v_{n',n,k}^{(l)} \geq 0$ are constant weights that we define as $v_{n',n,k}^{(l)} = v_{norm} \exp\left(-\frac{|r_{n,k}^{(l)} - r_{n',k}^{(l)}|}{2\eta l}\right)$, with $v_{norm}$ being a normalization constant ensuring $\sum_{l,n'} v_{n,n',k}^{(l)} = 1$, and $\eta^{(l)}$ being a scale-dependent constant. This latent variable, *M*, will serve as the abundance estimate for each element. Although it is not a conjugate prior, it will lead to non-negative analytical estimates for *M* and *R*. The variance parameters $\psi_{n,k}^2$ are assumed independent and assigned a conjugate inverse gamma distribution:

$$f(\psi) = \prod_{k=1}^{K} \prod_{n=1}^{N} \mathcal{IG}(\psi_{n,k}^2; \alpha_r, \beta_r) \qquad (5)$$

where $\alpha_r, \beta_r$ are positive free hyperparameters. They could be used to add prior knowledge of the sample in terms of average composition and variance, but they are by default set to zero to obtain a non-informative prior, which will be the case for all the results presented.

At this point the joint posterior distribution of this Bayesian model can be computed from the following hierarchical structure (dropping indices for clarity):

$$f(M, R, \psi | Y) \propto f(Y|R) f(R, M | \psi, V) f(\psi) \qquad (6)$$

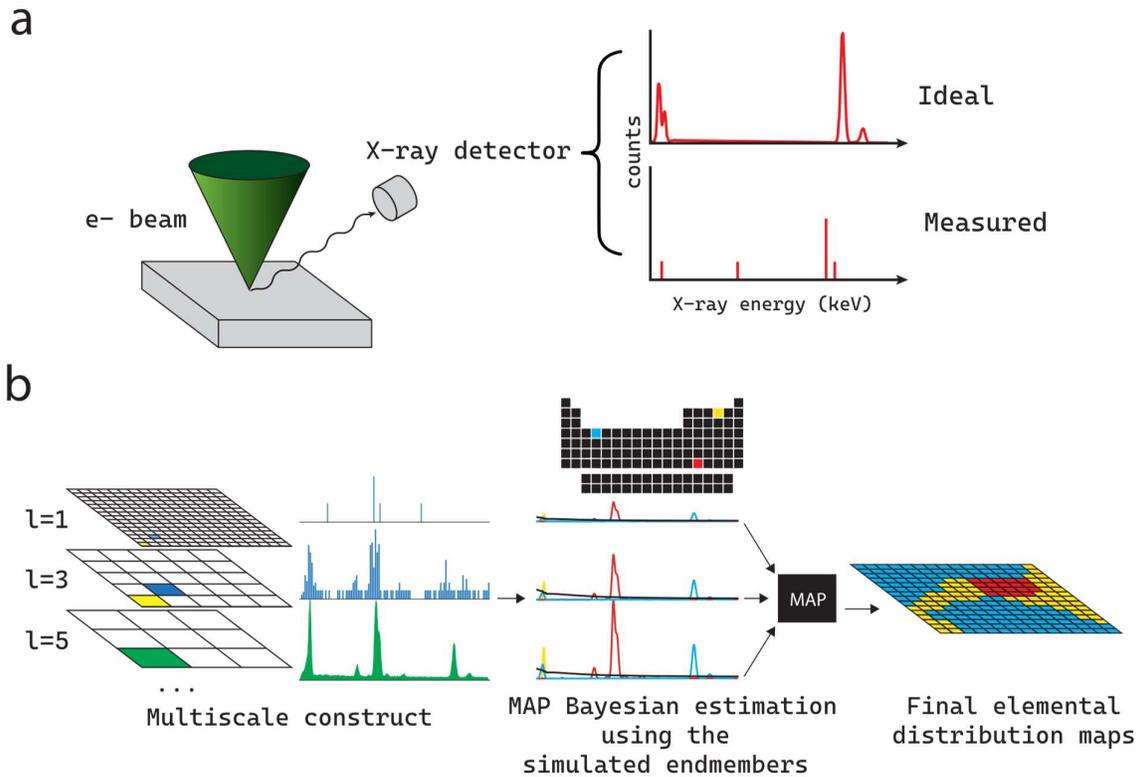

*Figure 1. The EDX challenge and robust multiscale Bayesian algorithm workflow.* *A) EDX is a very dose-inefficient technique. Commonly, the "measured" spectrum samples the "ideal", noise-free spectrum very sparsely. B) The RMB workflow proceeds as follows: first, the multiscale construct is computed by convolving the experimental spectrum image with spatial kernels of different sizes. Then, an initial estimation of the abundance for each element is obtained via the Sunsal algorithm[24], using endmembers corresponding to each element calculated with ESPM, which accounts for the relevant experimental parameters in the X-ray spectra acquisition. Then, we use MAP estimation for the abundances and variances within this observation model. At this point, the elemental abundance maps and their uncertainty are obtained.*

Where $f(Y|R)$ is given in (3), $f(R, M|\psi, V)$ in (4), $f(\psi)$ in (5). This distribution contains complete information regarding the parameters of interest (*R, M*) and their uncertainties *Ψ*. Here we consider a maximum a posteriori estimator (MAP) to obtain all the parameters. It should be noted that the abundance parameters *R, M, Ψ* are independent, allowing parallel optimization. The maximum a posteriori distribution is approximated using a coordinate descent algorithm[32,33] that sequentially maximizes the conditional distributions associated with each parameter until convergence to a local minimum of the negative log-posterior. The complete sequence of the algorithm with the analytical expressions for each iteration are detailed in the supplementary information. Its workflow is summarized in figure 1B.

*Data simulation*

To test our algorithm, a 90 x 90-pixel spectrum image consisting of a layer of $PbTiO_3$ (30 pixels wide) sandwiched between layers of $SrRuO_3$ (each 30 pixels wide) was simulated using the ESPM package.[19] This configuration, although not representative of the experimental sample (which only has bottom $SrRuO_3$ electrode), allows for an easy quantification of the resolution by estimating the width of the $PbTiO_3$ layer. The spectra contained 1024 energy channels between 0 and 20 keV. Spatially uniform blocks were used for both the $PbTiO_3$ and $SrRuO_3$ (i.e. non-atomic resolution). Different virtual acquisition times were simulated, leading to different amounts of signal from 5 to 500 counts per pixel, as can be seen in the figure S1 in the supplementary information.

*Experimental data acquisition*

A 28 nm thick, high quality epitaxial $PbTiO_3$ thin film and a 22 nm thick bottom $SrRuO_3$ electrode were grown on a (110)o-oriented $DyScO_3$ substrate. At this film thickness and electrostatic boundary conditions, and under the epitaxial strain imposed by the substrate, $PbTiO_3$ is in the ***a/c*** phase, with domains where the polarisation axis is oriented out-of-plane alternating with domains where the polarisation axis is in-plane[34]. To perform the STEM measurements, a focused ion beam lamella was prepared from the sample by cutting along the (001)o of the substrate, using a ZEISS CrossBeam 540.

The STEM-EDX measurements were performed in a double-aberration corrected FEI (Thermo Fisher Scientific) Titan Themis 60-300 microscope equipped with a Super-X G2 EDX detector and operated at 300 kV high tension. The probe was set with current of 90 pA and a convergence angle of 20 mrad. The acquired SI was obtained over a 347 x 389 pixel scan, with a pixel size of 12.68 pm and dwell time of 20 µs. The EDX spectra were acquired over the 0–20 keV energy range. The SI was acquired for 10 minutes, with individual frames acquired every 3.5 seconds. Beam damage started to be observable in the last few frames of the acquisition due to the accumulated dose. In order to avoid these beam damage effects, the data reported here correspond to the integrated counts from up to the first 350 s/100 frames of the acquisition. The total collected signal amounts to an average of 200 counts per pixel under these conditions.

*Sample growth*

The PbTiO$_3$/SrRuO$_3$ heterostructure was deposited using our in-house constructed off-axis radio-frequency magnetron sputtering system. The SrRuO$_3$ electrode was deposited on the (110)o-oriented DyScO$_3$ substrate from a stoichiometric target at 660 °C in 100 mTorr of O$_2$/Ar mixture of ratio 4:80, at a power of 80 W. The PbTiO$_3$ thin film was then deposited at 560 °C, in 180 mTorr of a 20:29 O$_2$/Ar mixture, at a power of 60 W, and using a Pb1.1TiO$_3$ target with 10% excess of Pb to compensate for its volatility. Huettinger PFG 300 RF power supplies were used in power control mode. The sample holder was grounded during deposition, but the sample surface was left floating. Atomic force microscopy measurements, X-ray diffraction measurements, and vertical piezoresponse force microscopy measurements are available in the supplementary information, which demonstrate the film quality and its microscale ferroelectric configuration.

## Results

*Simulated Data:*

In this section, we analyse the simulated SIs, which have a known ground truth, with our RMB algorithm. The output of the RMB algorithm is the signal of the selected elements present in a given SI ($r_{n,k}$), as well as the uncertainty of this measurement ($\Psi_{n,k}$). The maps $r_{n,k}$ are the weight of the atomic X-ray emission probability ($s_{k,t}$) in the model. Therefore, they allow direct quantification of the atomic percentage of each element after normalizing by $\sum_k r_{n,k}$.

Here, we focus on the quantification accuracy of the algorithm under different noise conditions. The simulated SI contains a layer of PbTiO$_3$ sandwiched between two layers of SrRuO$_3$ (see figure S1). For this sample three synthetic SIs were simulated, each having different virtual acquisition times that were adjusted to achieve an average detection of, respectively 5, 50, and 500 X-ray counts per pixel.

Figure 2 compares results from applying the RMB method to the synthetic datasets to results obtained using a standard quantification workflow, focusing on the elements Sr and Ti. (The quantification of all elements in the sample can be found in supplementary information, figures S2-S6). In the RMB approach, we estimate both the abundance of each element and the background at the same time, using the atomic X-ray emission spectra and the estimated Bremsstrahlung background modelled by ESPM[19]. Then, we calculate each atomic percentage by dividing the elemental abundance by the sum of all elemental abundances. The RMB algorithm requires the definition of a multiscale downsampling strategy, for which a single kernel of 3 x 3 pixels was chosen in this case. On the other hand, the standard quantification method consists of integrating the relevant spectral line intensities (in this case Ti K$_α$ and Sr K$_α$) after background fitting;

these intensities are then quantified using the Cliff-Lorimer (CL) equations[2]. This procedure was performed in using the relevant routines implemented in HyperSpy[35]. To provide a fair comparison with the RMB method, the signal was convolved with a 3 x 3-pixel kernel prior to making this standard quantification.

For both the standard and the RMB method, the quantification converges to the expected 20 at.% Ti in $PbTiO_3$ and 20 at.% Sr in $SrRuO_3$. As the signal increases, the noise in the quantification maps is reduced, and the quantification histograms narrow, for both methods. Critically, however, the maps from the RMB approach consistently show much less noise than those from the standard method. Correlated to this, the RMB gives per pixel at.% histograms that consistently have narrower distributions that those for the standard method, indicating that the RMB approach is more precise at all given SNR.

The RMB requires a downsampling strategy which, to some extent, trades a signal increase/denoising for a lower spatial resolution. To investigate this trade-off, we compare the outcome for a range of downsampling strategies. These results are contrasted against a "standard" noise reduction approach, whereby the SI is convolved with a kernel followed by making a CL quantification. The kernel size for the "standard" approach was varied from 3 x 3 to 11 x 11 pixels. The RMB method, on the other hand, applies multiscale downsampling. For it we used sets of downsampling factors {1, 3, 5, ..., 2$l$+1}. In our analysis, we compare the RMB approach using a downsampling set of {1...,2$l$+1} against the standard method using a 2$l$+1 kernel size, thereby ensuring an equivalence in the spatial origin of the signal for both cases. Hereafter, the maximum 2$l$+1 kernel size of a RMB analysis is simply referred to as its kernel size.

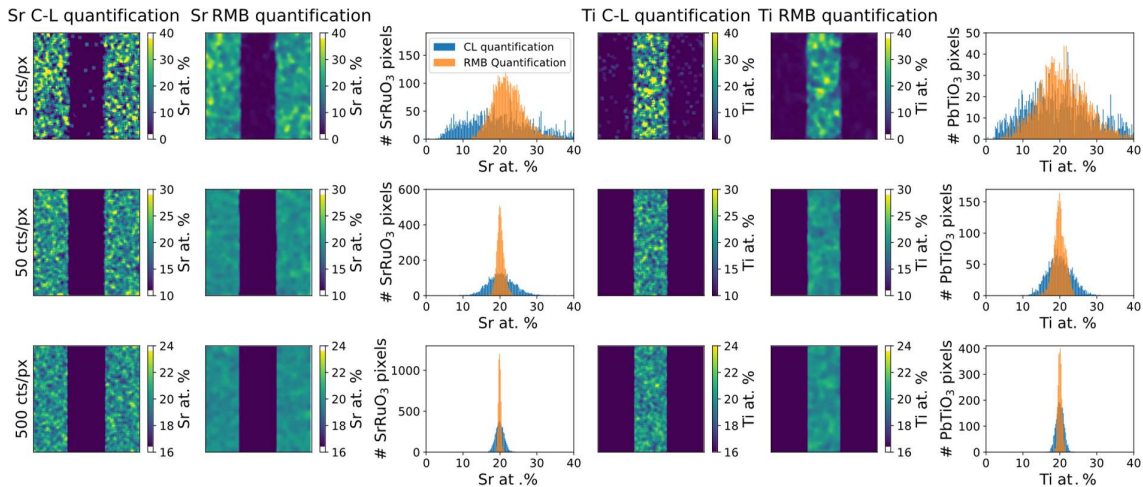

*Figure 2. Quantification accuracy. In the first column, Sr at.% maps calculated through the standard quantification method of spectral line integration followed by CL quantification are shown. The second column of panels shows the Sr at.% map calculated through RMB. The third column shows the pixel histogram of the obtained Sr maps. Columns four to six show the equivalent results for Ti. Each row corresponds to the results for a spectrum image with 5, 50, 500 X-ray counts per pixel, respectively.*

Figure 3 shows the results of these comparisons for kernel sizes 3, 7 and 11, this time comparing Pb maps as opposed to Sr and Ti for figure 2. The results for all elements can be found in the supplementary information, as are the full results for kernel sizes 5 and 9. Since our simulated map contains only two distinct, uniform phases, it is reasonable to assume that the classic Otsu thresholding algorithm[36,37] applied to the maps should recover the ground truth distribution of the phases. After performing said thresholding, the misclassified pixels that were identified for each method are also shown in figure 3.

The results in figure 3 highlight that, while loss of spatial resolution can be seen in both methods as kernel size increases, for each kernel setting this loss is much less pronounced when using the RMB algorithm than when using the conventional processing. Moreover, the SNR boosting in the Pb maps is much more effective for the RMB approach. This can be quantified by measuring the $SNR \equiv \frac{\sigma_{Pb}}{I_{pb}}$ (i.e. the standard deviation for the signal of a given element in a uniform region divided by the intensity of this signal) on the simulated $PbTiO_3$ phase against loss of resolution. We quantify the latter property by measuring the average width of the $PbTiO_3$ phase after thresholding and dividing it by the ground truth width, thus obtaining the feature broadening ratio associated to each kernel size for each method. These values are plotted in figure 4.

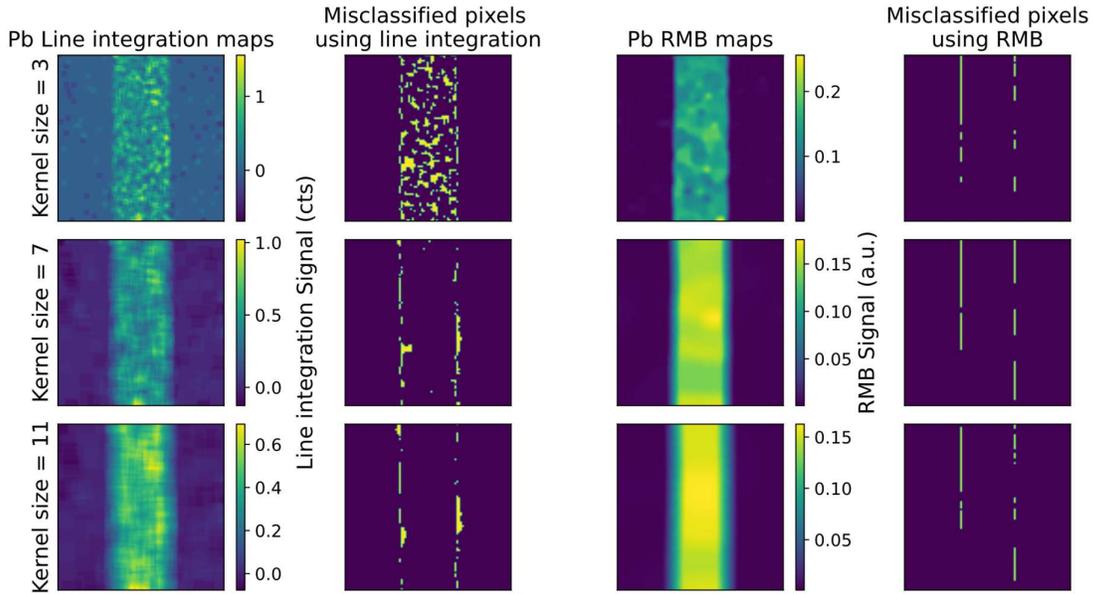

*Figure 3. Kernel size effect.* The first columns show the Pb signal maps obtained from the 5 cts/px SI after convolving the signal with uniform kernels of different sizes, followed by background subtraction and Pb $L_\alpha$ line integration. The second column corresponds to the misclassified pixels after applying Otsu thresholding to these maps and comparing with the ground truth (yellow for misclassified pixels). The third column corresponds to the Pb maps obtained with the RMB algorithm, with each row using the kernel size of the left column as its maximum kernel size in the multiscale downsampling. The fourth column corresponds to the misclassified pixels after applying Otsu thresholding to Pb RMB maps.

The SNR not only is higher on the RMB maps, but also increases at a much higher rate (note the logarithmic scale for the SNR in figure 4) when increasing the kernel size, while keeping a better spatial resolution.

As well as producing quantified maps, the RMB algorithm directly quantifies the uncertainty of the measurement. Figure S11 shows the evolution of the standard error, derived from the square root of the variances $\psi_{n,k}$, as the kernel size increases. There, it can be seen that, although the overall uncertainty decreases at higher kernel size, it increases at the interface between the two phases. This is intuitive as larger kernels mix the signal of both phases. Therefore, if one seeks a precise quantification of the bulk of a phase, higher kernel sizes should be used, while if the goal of an experiment is the precise location of the interface between two phases, the maximum kernel size on the multiscale strategy should be kept small. We further note that the uncertainty maps are useful for gauging the reliability of the analysis at each position.

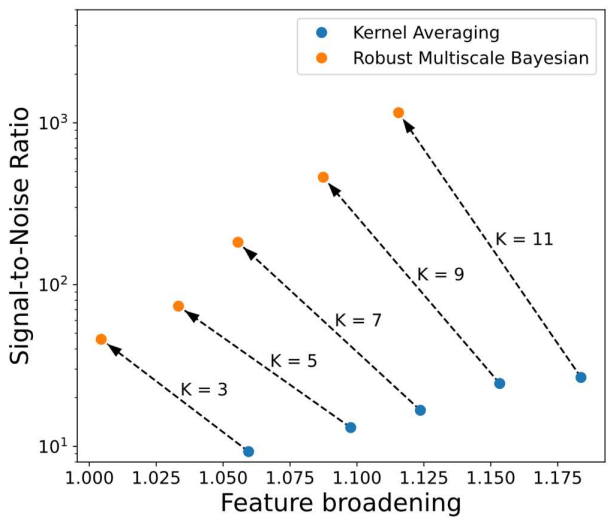

**Figure 4. SNR vs resolution evaluation.** The measured SNR plotted against feature broadening ratio when applying kernels of different sizes (K) on the 5 X-ray counts/pixel SI, for simple kernel averaging and for the RMB algorithm.

*Experimental Data*

In this section, we analyse an atomic-resolution STEM-EDX SI acquired from a $PbTiO_3$ film, using both standard and RMB methodologies. While the elemental concentrations obtained from atomic-resolution EDX datasets are not always directly meaningful, owing to electron channelling and dechannelling processes[38], the atomic column spatial features provide a useful means of evaluating the spatial quality of the processed maps. Moreover, such maps contain strong spatial–spectral correlations of the type that are implicitly unaccounted for using decomposition methods such as MVA or NMF. Finally, such a dataset also represents the case where beam-induced damage limits the SNR obtainable for the raw SI.

The SI is first analysed with the commercial Velox software from Thermo Fisher Scientific. For this purpose, a prefiltering of 5 pixels was used prior to obtaining the chemical maps of both Pb and Ti, which convolves the SI with a square kernel of said size. The net signal maps for each element were calculated after integrating different numbers of frames from the acquisition, which translates to different effective acquisition times. The results of this analysis are shown in the first row of figure 5. One way of assessing the spatial definition of a map is to calculate its fast Fourier transform (FFT), which displays dots that correspond to the spatial frequencies present in the map. In figure 5, FFTs of the Pb maps for acquisition times of 70 and 350 s are shown.

The second row shows the same dataset analysed with the RMB algorithm, using the same sets of integrated frames for comparison. The multiscale neighbour graph was set with sizes {1, 3 and 5}. Visually, the elemental maps show much sharper atomic columns when generated with the RMB algorithm. This improved spatial definition is directly seen in the Pb map FFTs. First, the FFT of the RMB Pb map at 70 s of acquisition time shows (110) spots (highlighted in blue in figure 5), which were lacking in the FFT from the maps obtained with the Velox software. Moreover, at 350 s, the (200) spots (highlighted in red in figure 5) are clearly observable in the FFT of the RMB maps, while they are lacking for the standard quantification approach.

This improvement in spatial resolution can be critical in specific applications. Here, we shall apply it to map the tetragonality of the $PbTiO_3$ unit cell. Moreover, as shown with ab-initio calculations by Bousquet et al.[40], for uniaxial ferroelectric distortions along the *c*-axis orientation, with $PbTiO_3$ in the P4mm phase, and relaxing the atomistic simulations performed at constant volume, the Ti displacement vector with respect to the centrosymmetric position defined by the surrounding Pb atoms is oriented in opposite direction and proportional to the polarisation[41]. Therefore, through a precise location of Pb and Ti atoms, it is also possible to extract an estimate of the local polarisation within each unit cell[42].

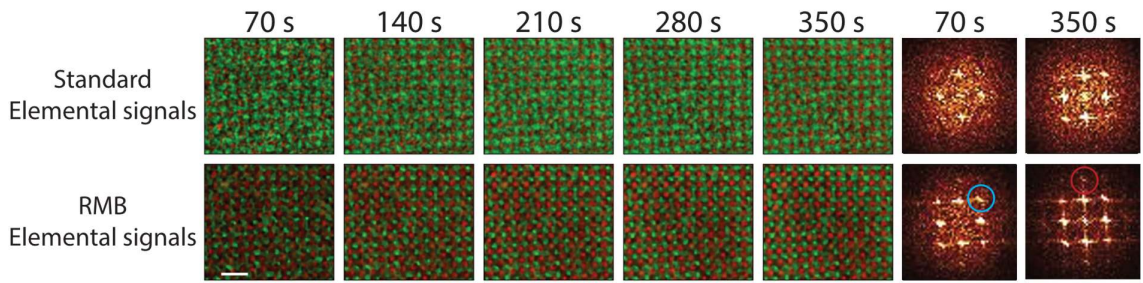

*Figure 5. Experimental EDX SI analysis.* *The Pb (red) and Ti (green) maps have been obtained with standard approach (using Velox software) and the RMB approach, for different effective acquisition times. The scale bar corresponds to 1 nm. The FFT of maps at 70 s and 350 s where calculated using Velox and the SciPy library[39] for the "standard" and RMB generated maps, respectively. The (110) spots (blue) and (200) spots (red) for a cubic perovskite structure are highlighted.*

Here, we attempt to track the positions of the Pb atomic columns from the EDX maps using the Atomap library.[43] The atomic column positions were also located in the co-acquired high-angle annular dark-field (HAADF) STEM image. This image has a much higher SNR, and the Pb columns are very easy to identify because of their brighter contrast, since the HAADF signal intensity is approximately proportional to $Z^{1.6-1.9}$.[44] This allows the positions obtained in the HAADF image to be used as a ground truth, enabling the quantification of the error in the positions obtained from the noisier EDX maps. Figure 6A shows the HAADF image and the located Pb atomic columns. Figures 6B and 6C show the Pb maps corresponding to an acquisition time of 350 s obtained through the standard (Velox) and the RMB approaches, respectively. The same exact parameters were used in Atomap to locate the atomic columns presented in Figures 6A–C: in brief, a minimum pixel distance of 15 was chosen for an initial column location, followed by 2D gaussian refinement. When using the Pb EDX map obtained through the standard approach, a number of columns are incorrectly detected, due to its poor SNR. In contrast, all the atomic columns are correctly identified in the RMB map correctly. The precision of each analysis is quantified by calculating the mean distance between each atomic position obtained from the HAADF map, and the closest atomic position obtained from the maps in figures 6B, C. This mean error was calculated for the different acquisition times considered in figure 5, obtaining the results of figure 6D, which demonstrate a consistent improvement in atomic column spatial localization when using maps generated with the RMB algorithm.

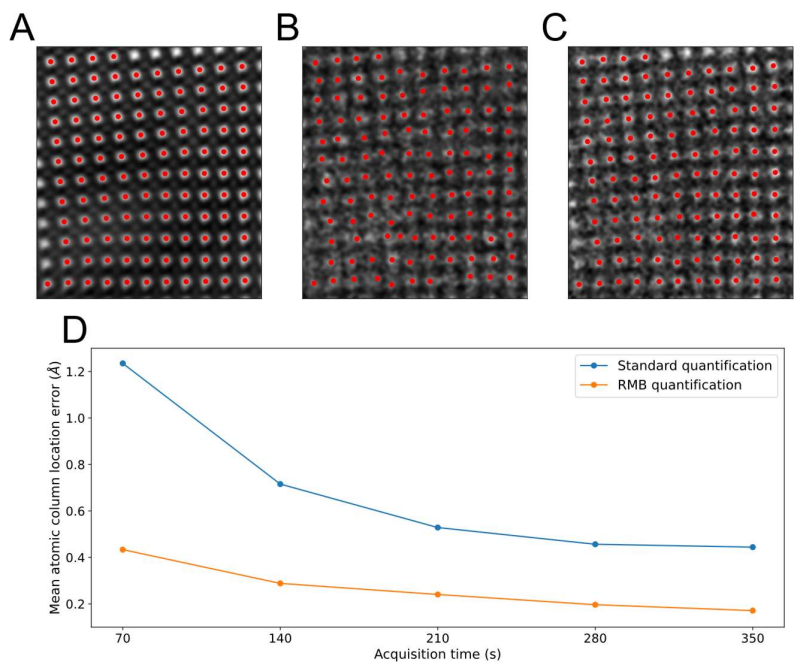

*Figure 6. Atomic column location measurements. HAADF image (a) and Pb maps obtained through standard (b) and RMB quantifications (c). The atomic column positions obtained with atom map are plotted with red dots in each image. d) Mean error in the atomic column position relative to the HAADF image for the standard and RMB quantification methods.*

Although this example is trivial, in the sense that the atoms can be located with ease in the HAADF image, it is easy to think of cases where atom location in EDX maps is desirable. For example, if the elements of interest have a similar atomic number, the HAADF image is no longer suitable for this purpose, and atomic-resolution EDX would be a solution. It should be noted that the precise tracking of atoms at this scale allows the measurement of relevant properties of the sample. In this case, the SI was acquired across a domain boundary between regions with ferroelectric polarization pointing along perpendicular directions: the upper-left side of the maps shown in Figure 6(A–C) display the position of the Pb atoms forming unit cells that are tilted and elongated in-plane (corresponding to the horizontal direction of the image, which we denote as the *a*-direction). This indicates that this region of the image corresponds to a domain with in-plane polarisation, labelled *a*-domain. In the lower-right side of the maps, the unit cells have a different tilt orientation and are elongated out-of-plane (coinciding with the vertical direction of the image, which we denote as the *c*-direction). This indicates that the region of the image corresponds to a domain with out-of-plane polarisation (*c*-domain). In the lower magnification, overview STEM-HAADF image shown in figure 7A, the domain boundary between these two regions appears with a brighter contrast.

It is possible to track the domain distribution at the unit-cell level by measuring the unit cell parameter along the vertical (*c*) and horizontal (*a*) directions in the image through the Atomap measurements. Unit-cells with out-of-plane polarization have a *c/a* ratio greater than 1, while in-plane polarization unit-cells have a *c/a* ratio smaller than 1.

The obtained *c/a* ratio of each unit cell using both the HAADF image and the RMB map is shown in figures 7B and C, respectively. The good match between the two and further proves that the RMB maps are of sufficient quality to obtain relevant structural properties of the sample. As can be inferred from figure 6B, this measurement was not possible using the standard EDX analysis method.

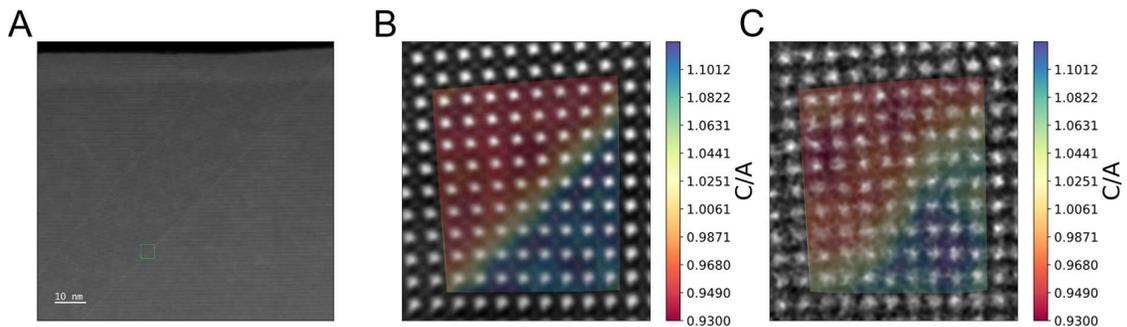

*Figure 7. Ferroelectric domain boundary measurement.* A) Overview HAADF image of the PbTiO$_3$ sample, with visible domain boundaries between regions with polarization in plane and out of plane. The EDX spectrum image was acquired from the region marked in green. B) HAADF image with an overlay of the corresponding *c/a* ratio for the central unit cells. C) RMB Pb map with an overlay of the corresponding *c/a* ratio for the central unit cells in the image. The two *c/a* ratio maps show good agreement with each other.

## Conclusions

In this work we have introduced a new algorithm that includes the physics of the X-ray emission process and exploits the spatial correlation between pixels to obtain improved atomically quantified elemental EDX maps. Our results on simulated data demonstrate that the algorithm improves SNR more efficiently than standard quantification approaches, while also maintaining a better spatial resolution than them. The quantification is also more chemically precise, since compositions for each pixel were closer to the ground truth at any noise level, compared to standard quantification. Moreover, thanks to the Bayesian framework, it provides a measurement of the quantification uncertainty for each pixel, essential for any quantitative measurement. Another potential benefit of this framework is that, once any initial probability distributions for the abundances are calculated, it is possible to easily update them via Bayes-like rules if additional signal is recorded. The adaptative update of the statistical model could allow on-the-fly estimation of abundances during an experimental STEM-EDX acquisition.

When applied to experimentally acquired data, for any given acquisition time, the RMB algorithm generated elemental maps that had higher spatial resolution and better SNR than the standard analysis methods. This allowed for the measurement of relevant structural properties of a material—in this case the tetragonality in $PbTiO_3$—with resolution at the atomic level.

It should be noted that, with signals so faint as the ones described in this work (as low as 5 total counts in a 1024 energy channel spectrum), the application of many matrix decomposition algorithms is not viable, as the data matrix is too sparse. In contrast, here we provide a universal denoising and quantification method, applicable to any EDX SI dataset. In addition, RMB does not introduce any of the biases that occur in matrix decomposition analyses of low-signal SIs. This more efficient processing of STEM-EDX data can translate to all the fields that already use the technique, especially in cases where the sample is beam sensitive or when short acquisition times are required.

## Data Availability

The simulated data can be easily reproduced using the ESPM library.[19] The experimental spectrum images are available from the corresponding author on reasonable request.

## Code availability

An implementation of the algorithm is available in GitHub[45].

https://github.com/PauTorru/RobustMultiscaleBayesian

## Acknowledgements

This work was supported by the UK Royal Academy of Engineering under the Research Fellowship Scheme (RF/201718/17128). We acknowledge the Interdisciplinary Centre for Electron Microscopy (CIME) at EPFL for providing access to their electron microscopy facilities. Lastly, we thank Prof. Jean-Marc Triscone and Prof. Patrycja Paruch from the University of Geneva for their continued support during the realization of this work.

## Author contributions

P.T. and A.H conceived the idea for this work. A.H. developed the mathematical models. P.T. and A.H developed the code and performed the analysis of the data. L.T. and C.L. synthesized the $PbTiO_3$ sample and performed the AFM, VPFM and XRD measurements. P.T. performed the STEM-EDX measurements. All authors contributed writing and editing the manuscript.

## Supplementary information

***Algorithm description:***

The estimation algorithm used is summarized in Algorithm 1.

| **Algorithm 1** Estimation algorithm: |
|---|
| 1: Input: |
| 2: **Y** , φ$^{1,\cdots,L}$ |
| 3: <u>Generate low resolution data:</u> |
| 4: Generate low-resolution histograms **Y** $^{(\ell)}$, $\ell \in \{1, \cdots, L\}$ using **φ**$^{1,\cdots,L}$ |
| 5: Estimate $\overline{r_{k,(l)}}$ , using Sunsal[24]. |
| 6: Compute **V** |
| 7: <u>Coordinate descent algorithm</u> |
| 8: while conv= 0 do |
| 9: Update **m** using analytical mode in (7) |
| 10: Update $r^{(\ell)}$ using analytical mode in (9) |
| 11: Update **Ψ** using analytical mode in (11) |
| 12: Set conv= 1 if the convergence criteria are satisfied |
| 13: end while |
| 14: <u>Output:</u> |
| 15: **M , Ψ** |

Followingly the analytical expressions used in each step of the algorithm:

1) Updating *M*: It is clear from Eq. (6) that the conditional distribution *M* results from Eq. (4). This is a normal distribution whose mean is given by

$$\widehat{m}_{n,k} = \frac{\Sigma_{l,n' \in v_n} v^{(l)}_{n',n,k} r^{(l)}_{n',k}}{\Sigma_{l,n' \in v_n} v^{(l)}_{n',n,k}} \qquad (7)$$

This equation highlights a weighted sum of the muti-scale abundance maps *r*.

2) Updating *R*: The parameters of *R* are independent, allowing parallel updating of $r_{n,k}^{(l)}, \forall n,k,l$. The conditional distribution of *R* is obtained by combining the likelihood in (3) and the prior in (4). Minimizing the negative-log of the conditional distribution reduces to:

$$\hat{r}_{n,k}^{(l)} = argmin_r \{r - \bar{s}_{n,k}^{(l)} \log(r) + \mathcal{H}(r)\} \qquad (8)$$

Where $\mathcal{H}(r) = \frac{1}{2\psi}(r - \mu_r)^2$ with $\psi_r^{-1} = \sum_{n'} \frac{v_{n',n,k}^{(l)}}{\psi_{n',k}}$ and $\mu_r = \sum_{n'} \frac{v_{n',n,k}^{(l)} m_{n',k}}{\psi_{n',k}}$. The minimum is analytically provided by[46]:

$$r_{n',k}^{(l)} = \frac{\mu_r - \psi_r + \sqrt{(\mu_r - \psi_r)^2 + 4\psi_r \bar{s}_{n,k}^{(l)}}}{2} \qquad (9)$$

3) Updating *Ψ*: The conditional distribution of the abundance variance $\psi_{n,k}$ is an inverse-gamma distribution given by

$$\psi_{n,k}|M,R,V \sim \mathcal{IG}\left[\frac{L+N}{2} + \alpha_r, \mathcal{K} + \beta_r\right] \qquad (10)$$

With $\mathcal{K} = \sum_{l,n' \in v_n} \frac{v_{n',n,k}^{(l)}\left(m_{n,k} - r_{n',k}^{(l)}\right)^2}{2}$. The mode is analytically given by

$$\hat{\psi}_{n,k} = \frac{\mathcal{K} + \beta_r}{\frac{L+\bar{N}}{2} + \alpha_r + 1} \qquad (11)$$

4) Stopping criteria: Two criteria are considered to stop the iterative coordinate descent algorithm for the abundance. The first is maximum number of iterations. The second evaluates the estimated parameter values and stops the algorithm if the relative difference between successive iterates is smaller than a threshold as detailed by Madsen et al.[47].

*Derivation of eq. (2)*

The likelihood in (1) can be written for the *n*th pixel as follows:

$$P(y_n|r_n) \propto \frac{\left[\sum_{k=1}^K r_{n,k}s_{k,t}\right]^{\sum_{t=1}^T y_{n,t}} \prod_{k=1}^K \exp\left[-\sum_{t=1}^T r_{n,k}s_{k,t}\right]}{\prod_t y_{(n,t)}!} \qquad (13)$$

Using Jensen's inequality on the log-likelihood, assuming $\sum_{t=1}^T y_{n,t} = \sum_{k=1}^K \bar{r}_{n,k}$, and only focussing on the abundance terms, leads to:

$$P(y_n|r_n) \geq \prod_k (r_{n,k})^{\bar{r}_{n,k}} \exp(-r_{n,k}S_k) \bar{Q}(y_{n,k}) \qquad (14)$$

We therefore adopt the approximation in (2) since the aim is to maximize the likelihood with respect to the abundances, which is also achieved by maximizing the right-side term.

*Data simulation*

Figure S1 shows the distribution of $SrRuO_3$ and $PbTiO_3$ in the simulated spectrum image as well as a representative spectrum in an individual pixel for the 5 different simulated virtual acquisition times:

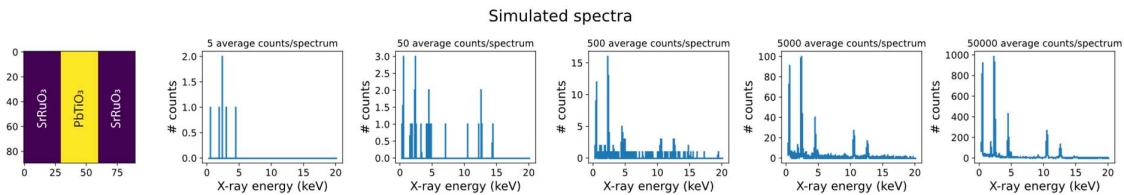

Figure S1. $PbTiO_3$ (in yellow) and $SrRuO_3$ (in dark purple) "ground truth" distribution and spectra contained in a single pixel for the different virtual acquisition times, that lead to different average X-ray counts per spectrum.

Below are the quantification results for all elements:

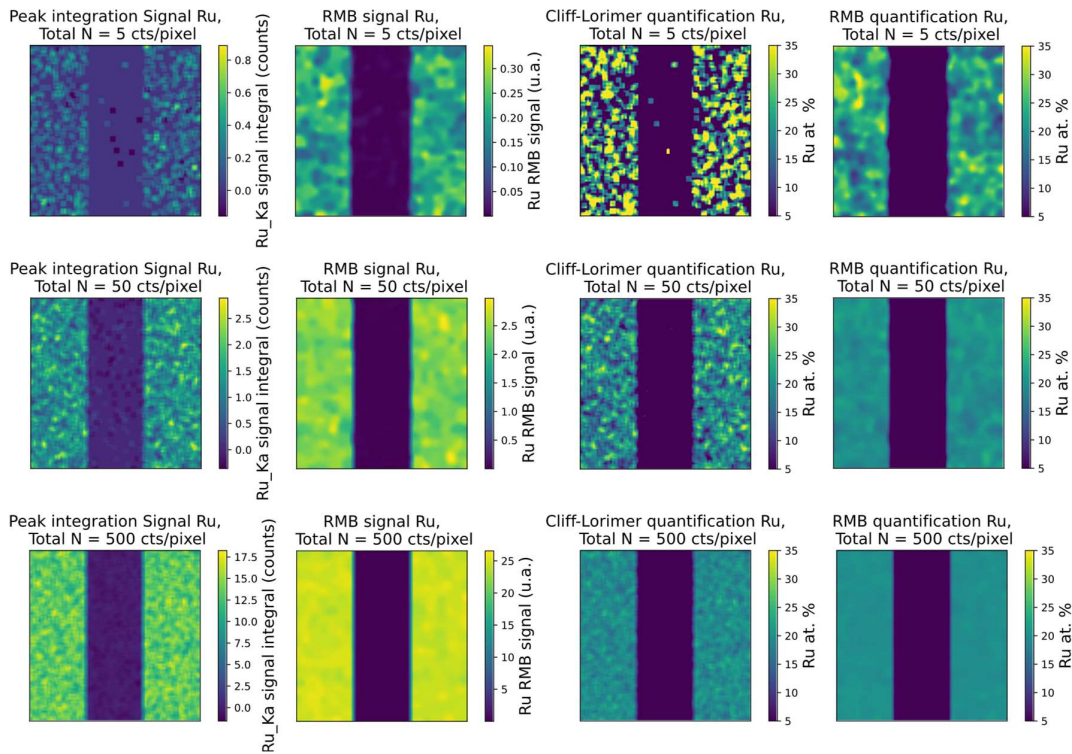

Figure S2. Standard and RMB quantification of Ru for the simulated datasets of 5, 50 and 500 cts/pixel.

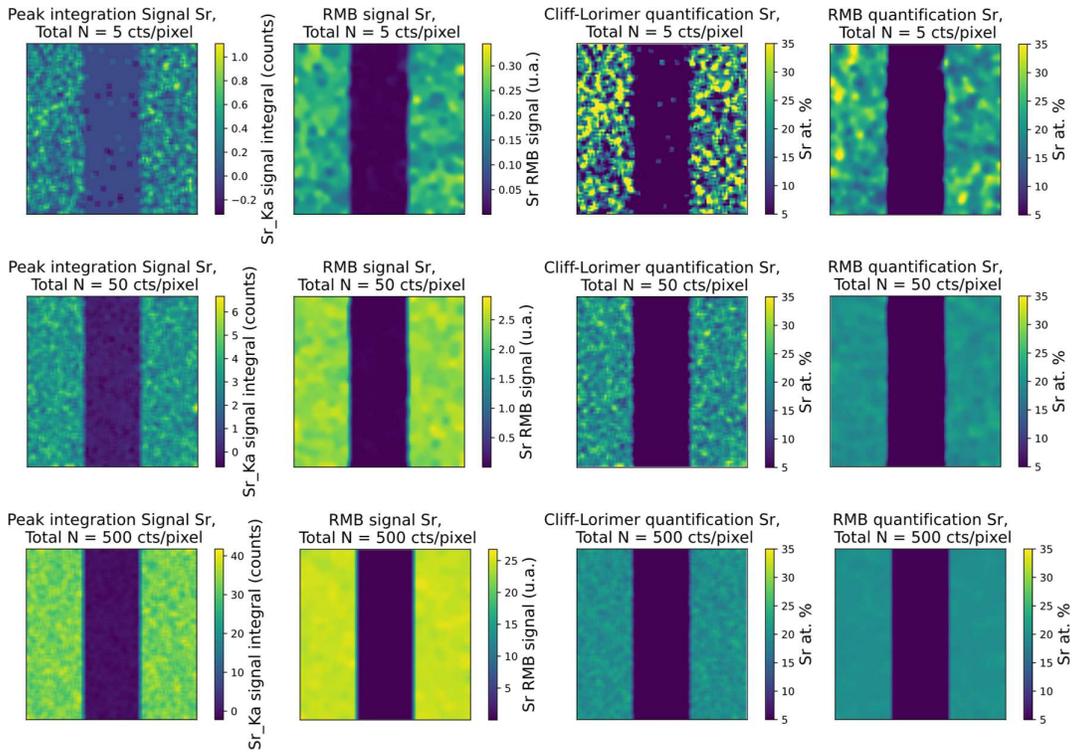

Figure S3. Standard and RMB quantification of Sr for the simulated datasets of 5, 50 and 500 cts/pixel.

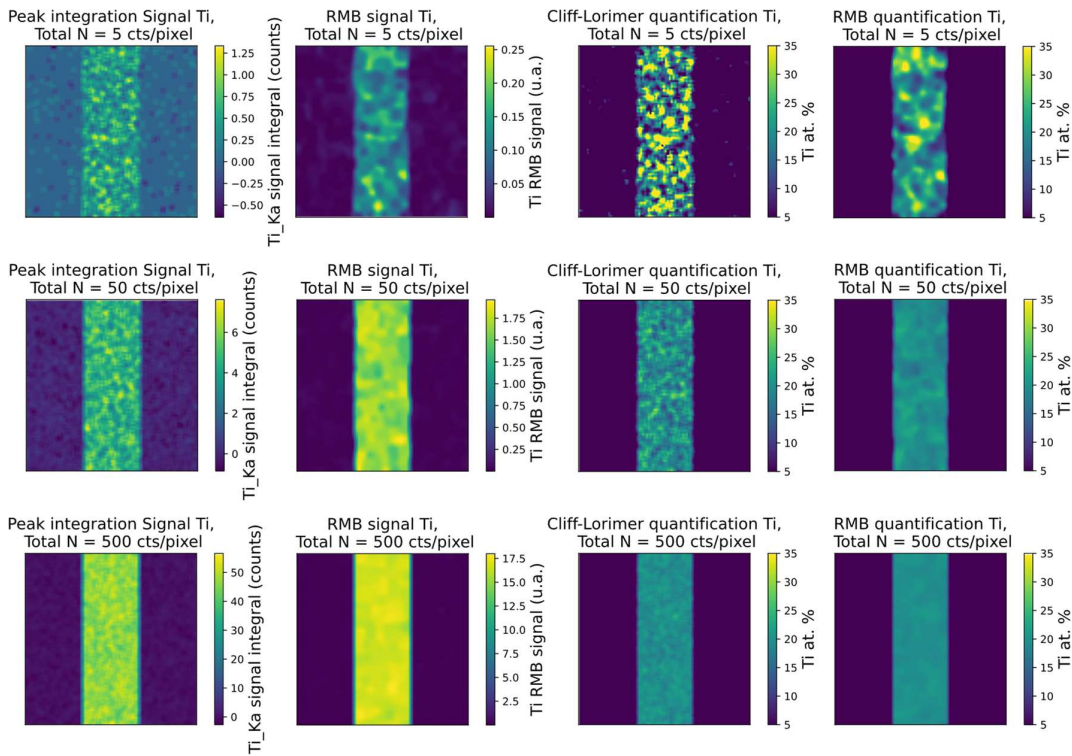

Figure S4. Standard and RMB quantification of Ti for the simulated datasets of 5, 50 and 500 cts/pixel.

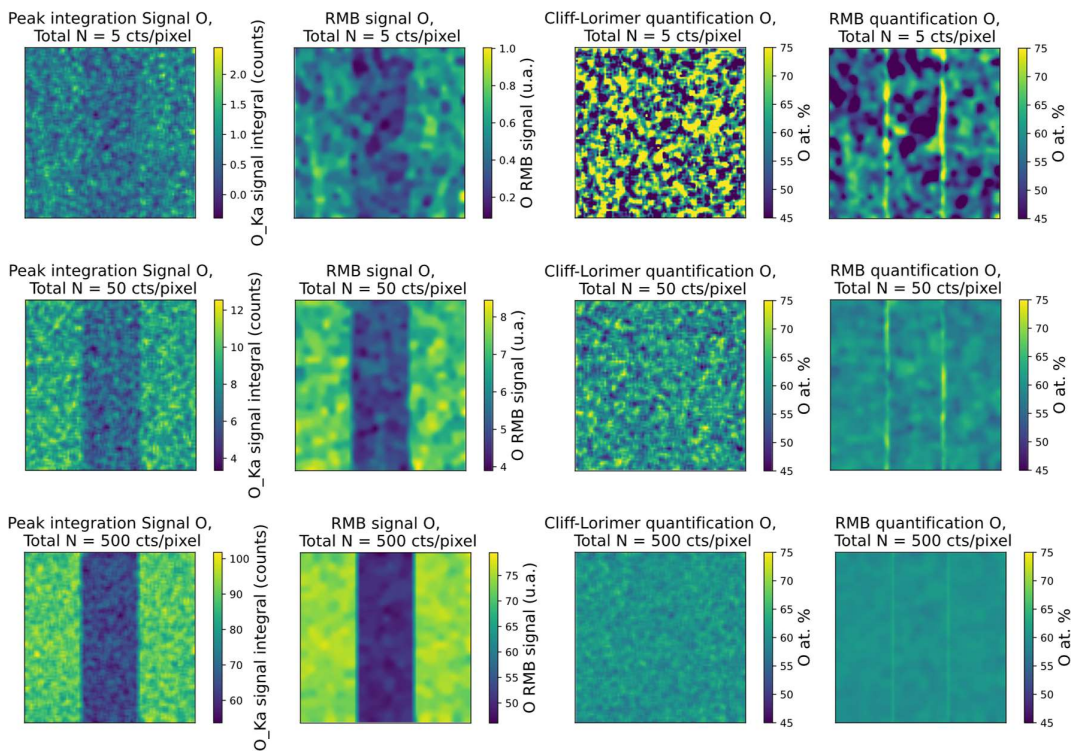

Figure S5. Standard and RMB quantification of O for the simulated datasets of 5, 50 and 500 cts/pixel.

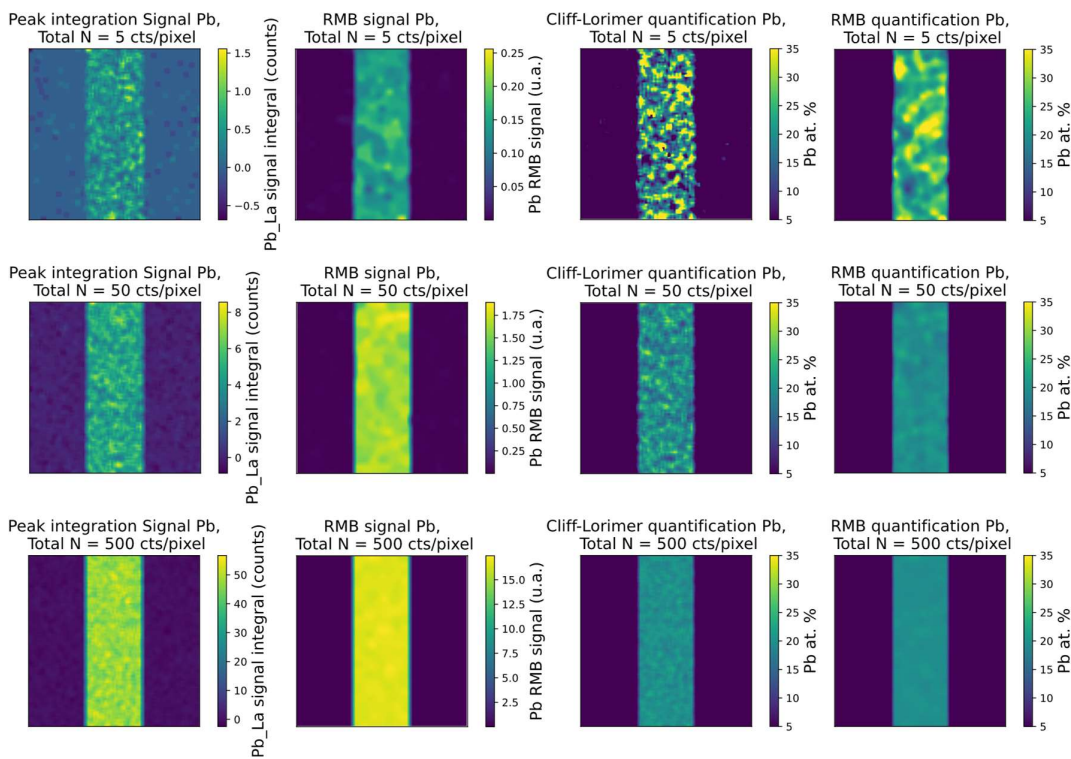

Figure S6. Standard and RMB quantification of Pb for the simulated datasets of 5, 50 and 500 cts/pixel.

Below, the results for the Otsu threshold labelling accuracy of Ru, Ti, Pb and Sr can be found (oxygen has a uniform distribution over the sample and therefore the thresholding accuracy does not apply):

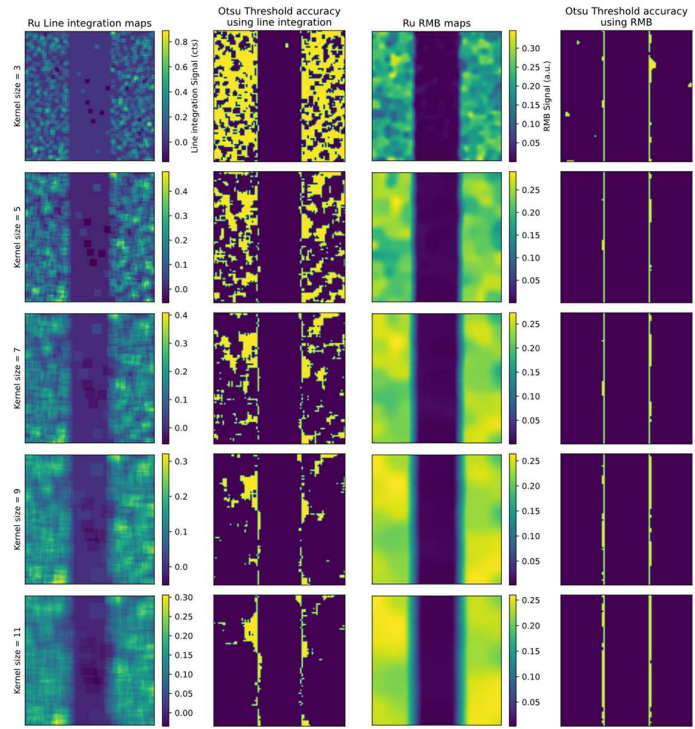

Figure S7. Kernel size effect for standard and RMB quantification of Ru.

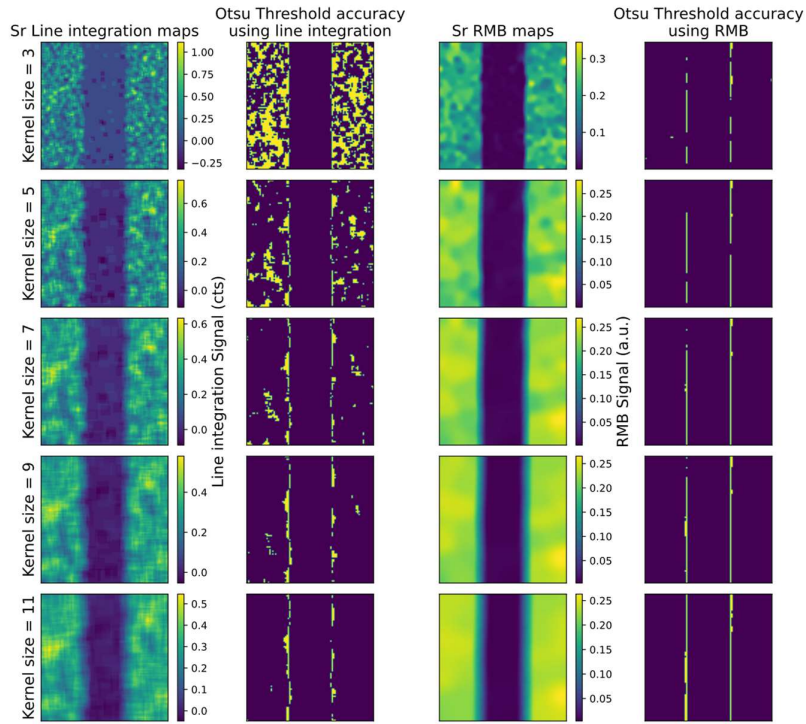

Figure S8. Kernel size effect for standard and RMB quantification of Sr.

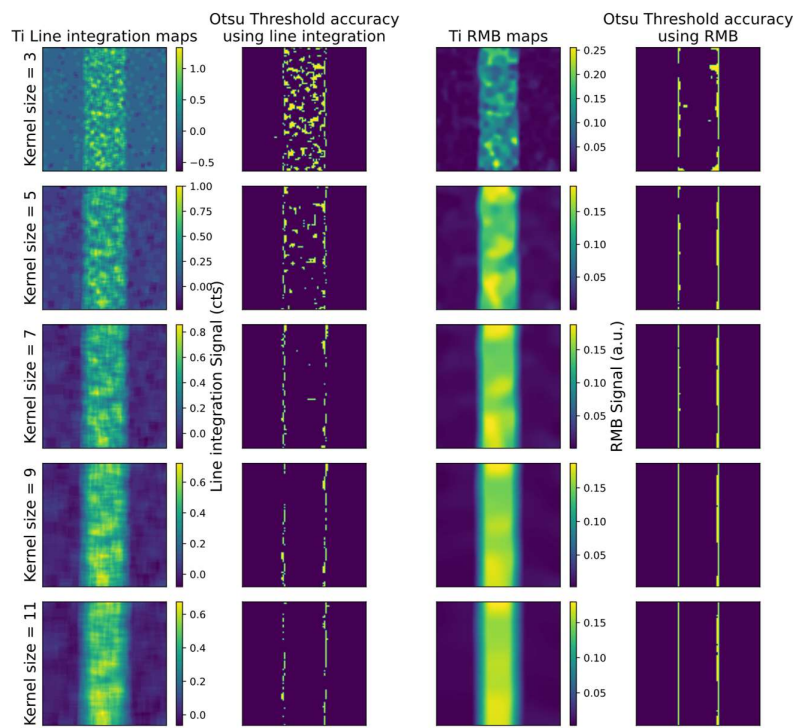

Figure S9. Kernel size effect for standard and RMB quantification of Ti.

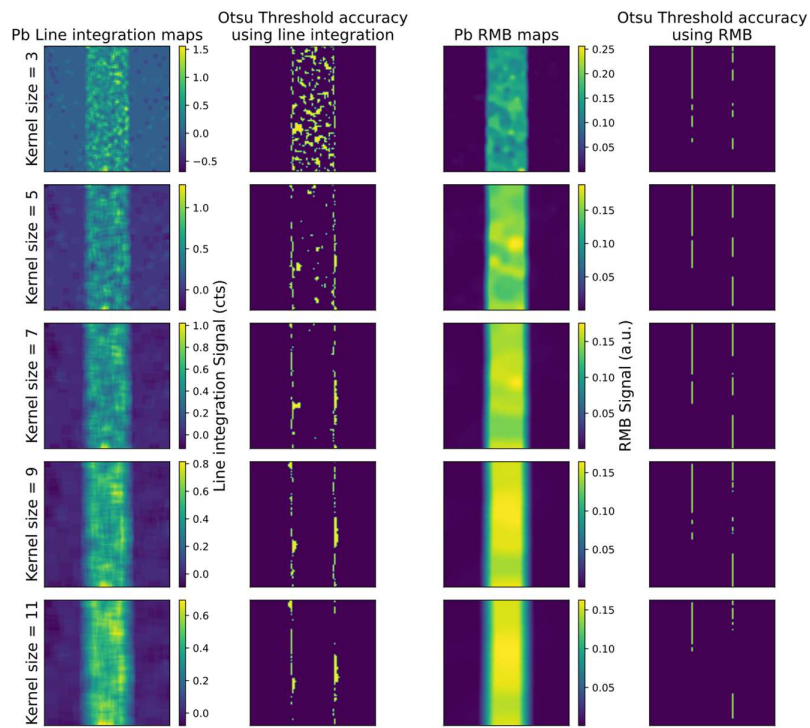

Figure S10. Kernel size effect for standard and RMB quantification of Pb.

*Uncertainty measurement.*

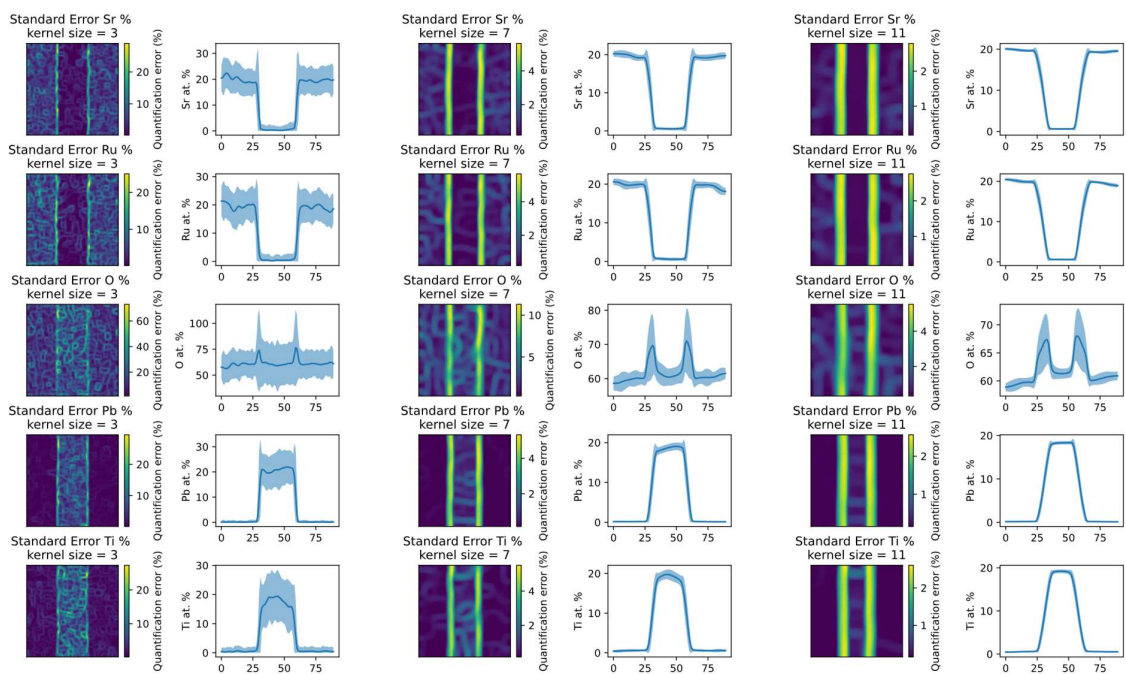

Figure S11. Evolution of the RMB-measured uncertainty as the kernel size increases. The profiles show the composition averaged along the vertical direction, and its standard error (in lighter blue).

*AFM XRD and PFM results*

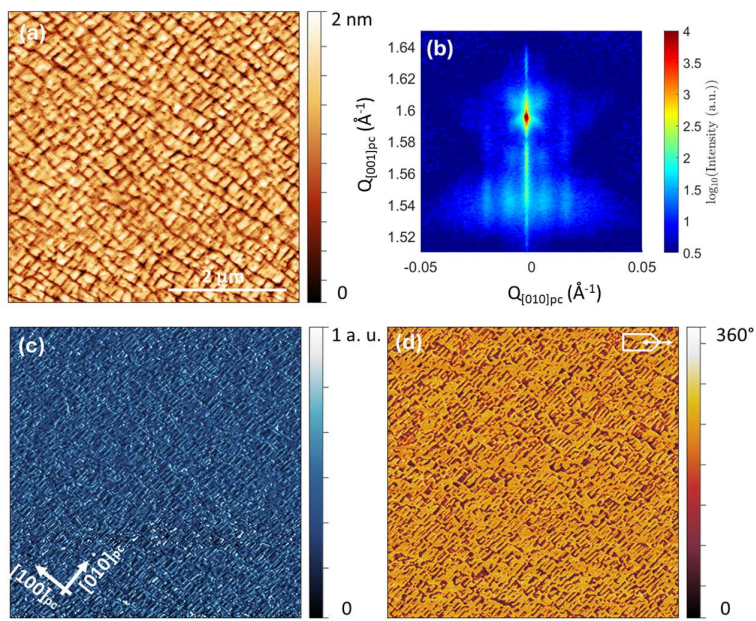

Figure S12: Characterization of the PbTiO$_3$ sample (a) 5 x 5 µm$^2$ atomic force microscopy topography image, the orientation is fixed with respect to the substrate pseudocubic axes

[100]$_{pc}$ and [010]$_{pc}$. (b) Reciprocal space map around the (001)$_{pc}$ peak of the substrate in the Q$_{[100]pc}$ – Q$_{[010]pc}$ plane. Vertical piezoresponse force microscopy (c) amplitude and (d) phase images. The cantilever orientation is sketched in the top right corner of the phase image (d).